\documentclass[final,3p,times,twocolumn]{elsarticle}
\def\ph2{{\it p}-H$_2$}
\def\od2{{\it o}-D$_2$}

\usepackage{amssymb}

\journal{Results in Physics}

\begin{document}

\begin{frontmatter}



\title{Computer simulations of supercooled liquid hydrogen mixtures and the possible crystallization slowdown}


\author[inst1]{Massimo Boninsegni}

\affiliation[inst1]{organization={Department of Physics},
            addressline={University of Alberta}, 
            city={Edmonton},
            postcode={T6G 2E1}, 
            state={Alberta},
            country={Canada}}



\begin{abstract}
Metastable liquid mixtures of parahydrogen and orthodeuterium are studied theoretically by means of computer simulations. No reduced  propensity of  the mixture to undergo crystallization is observed, compared to that of pure liquid parahydrogen. Demixing of the two species as a precursor of crystallization is not observed either. 

\end{abstract}




\end{frontmatter}


The physical behavior of moderately and highly quantal fluids cooled below their crystallization temperature is a fascinating subject, not only because it allows for the investigation of quantum-mechanical effects in structural phase transitions, but also because supercooled liquid substances could conceivably display physical properties that the equilibrium solid phase does not possess. This has been the main motivation underlying decades of experimental \cite{PhysRevB.24.467,PhysRevLett.56.2380,PhysRevB.44.9639,PhysRevB.53.11451} work aimed at detecting a  hypothetical  superfluid phase  of liquid parahydrogen (\ph2) cooled below its freezing temperature (at saturated vapor pressure), namely $T_F=13.8$ K.

Parahydrogen has a strong tendency to crystallize, even in confinement \cite{PhysRevB.93.104501,ultimo} and in reduced dimensions \cite{PhysRevLett.111.235303,PhysRevB.70.193411}, and there is now robust theoretical evidence that the superfluid transition of liquid \ph2 predicted decades ago \cite{ISI:A1972M646400017}  does not take place. Rather, the supercooled fluid remains in the normal phase, due to the strong suppression of quantum-mechanical exchanges caused by the relatively large diameter of the repulsive core of the intermolecular interaction, close to 3 \AA\  \cite{PhysRevB.97.054517}. Nevertheless, it remains an interesting theoretical question whether and how fluid \ph2 could be brought to temperatures significantly below $T_F$. 

It has been recently claimed \cite{PhysRevB.89.180201} that a significant   slowdown of crystallization occurs in mixtures of \ph2 and its heavier orthodeuterium (\od2) isotope, the effect generally more pronounced with increasing \od2 concentration $X$, at least up to $X\sim50\%$. Such a prediction is based on Raman spectroscopy of liquid microjets, supplemented by Quantum Monte Carlo (QMC) simulations of the mixtures at temperature as low as $T=13$ K, i.e., only slightly below $T_F$.  

We revisit here the simulation part of the argument provided in Ref. \cite{PhysRevB.89.180201}; specifically, we carried out our independent calculations, arriving at the opposite conclusion, namely that there is no structural evidence of suppressed crystallization in the mixtures.
The key difference between our simulation and theirs, besides the size of the system studied (1,024 particles, over three times greater), is that our  cooling protocol, which is the same as in Ref. \cite{PhysRevB.97.054517}, allows us to investigate  the behavior of the metastable fluid down to a much lower temperature, namely $T=2$ K.

The QMC methodology adopted here is the same as in Ref. \cite{PhysRevB.89.180201}, namely the canonical \cite{mezz,mezz2} continuous-space Worm Algorithm \cite{worm,worm2}, though it is worth noting that because quantum-mechanical exchanges of identical particles are virtually non-existent, details of the simulation are essentially the same as in Ref. \cite{sampling}. We utilized the standard  microscopic model of the system, based on the Silvera-Goldman pair potential to describe all intermolecular interactions (though not explicitly mentioned in Ref. \cite{PhysRevB.89.180201}, the same choice appears  to have been made therein). All of the results shown here are for a density $\rho=0.023$ \AA$^{-3}$, i.e., the freezing density of \ph2, and for a temperature $T=2$ K, the lowest considered here. It is worth mentioning that simulations carried out at temperature $T=4, 8$ and 10 K yield essentially the same results  shown below.

\begin{figure}[h]
\centering
\includegraphics[width=\linewidth]{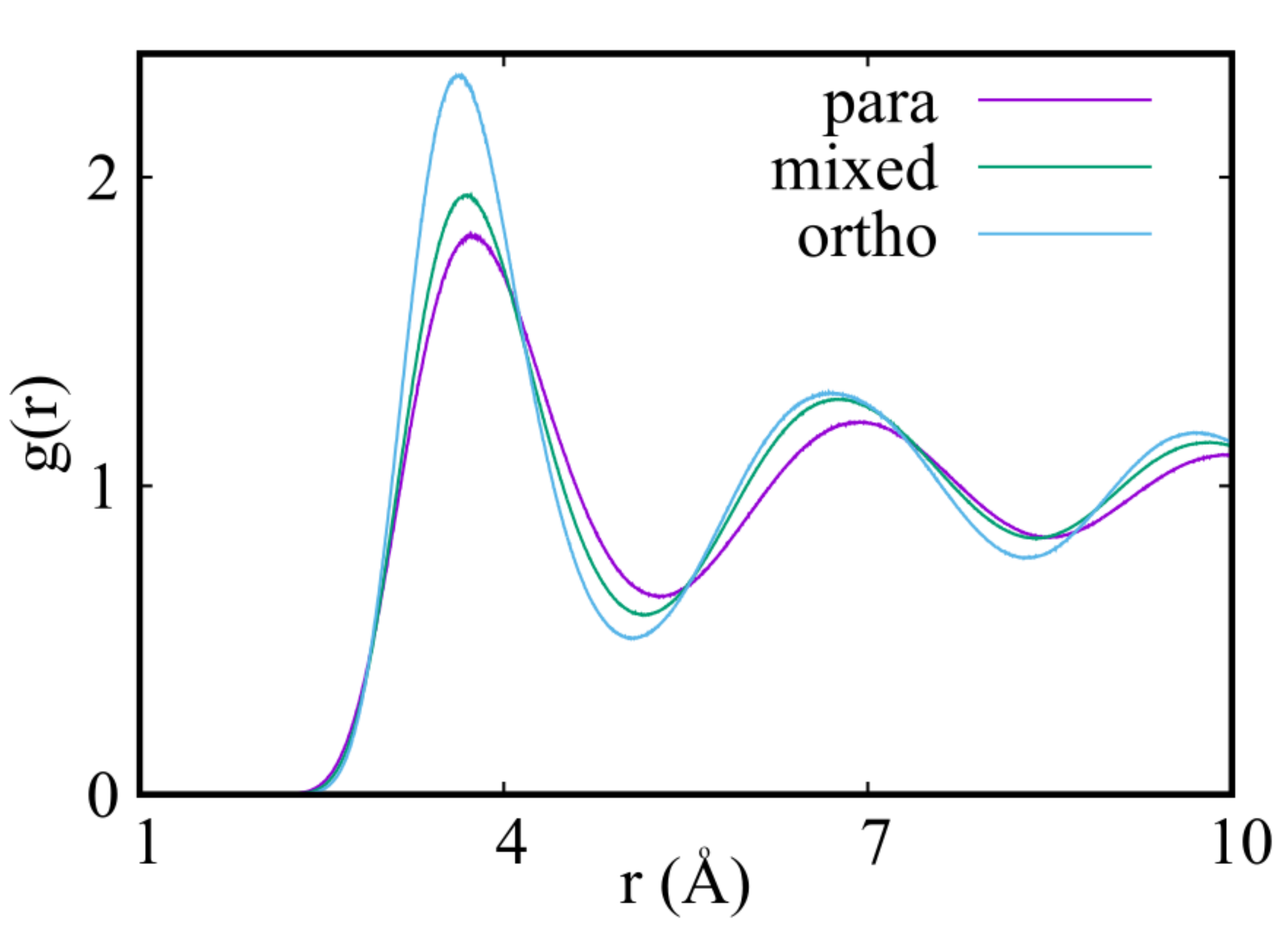}
\caption{Computed same-species and mixed pair correlation functions in a mixture with a $X=0.49$ \od2 concentration. }
\label{f0}
\end{figure}

\begin{figure}[h]
\centering
\includegraphics[width=\linewidth]{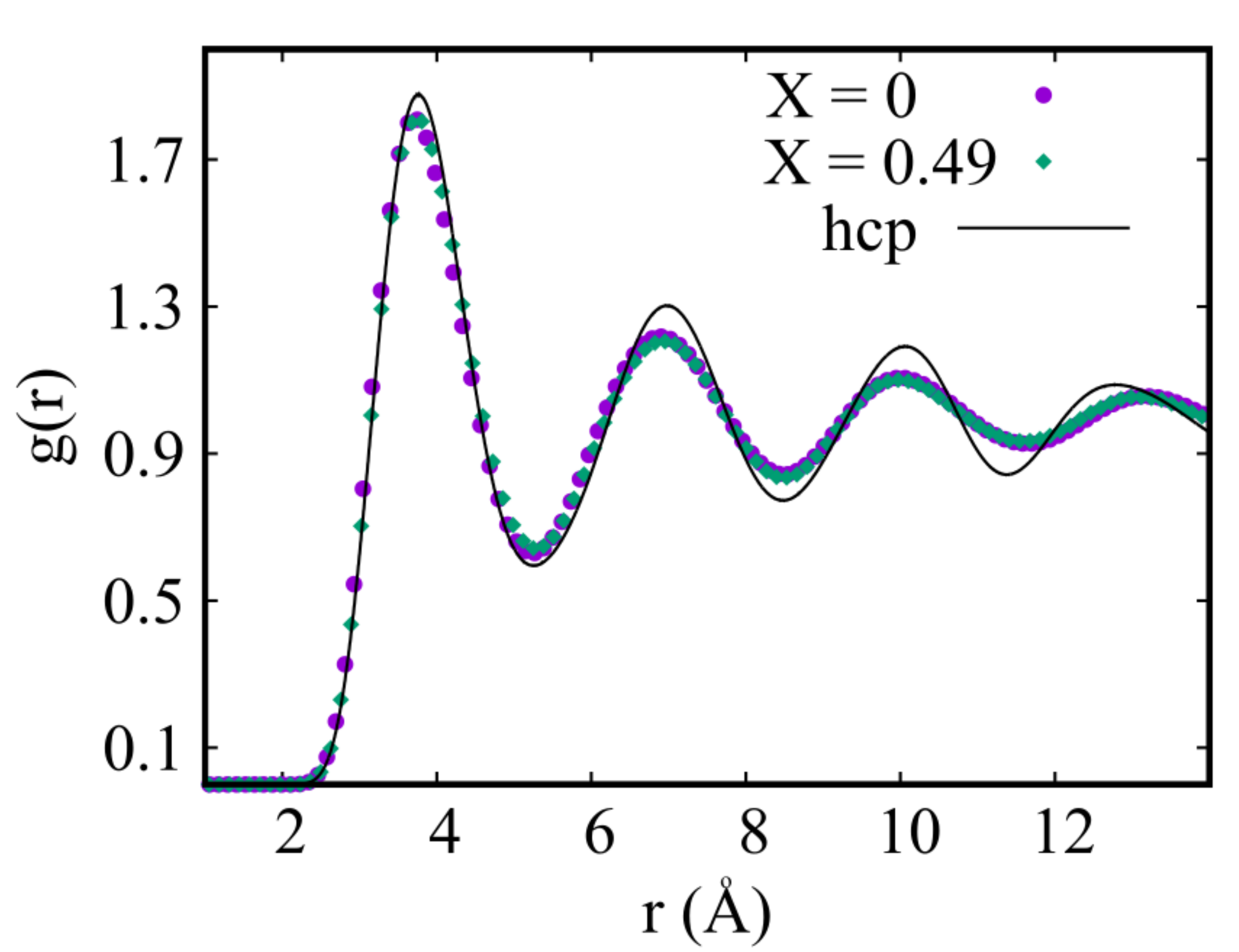}
\caption{Pair correlation function for metastable pure fluid \ph2 ($X=0$, circles), as well as in a mixture with a $X=0.49$ \od2 concentration (diamonds). Also shown (solid line) is the same quantity for the crystalline (hcp)  phase of \ph2.}
\label{f1}
\end{figure}

Fig. \ref{f0} shows the pair correlation functions for a mixture with \od2 concentration $X=0.49$. The result is entirely consistent with that shown in Ref. \cite{PhysRevB.89.180201}, in that the heights and the positions of the peak are noticeably different for the same-species and mixed correlation function, to indicate  significant quantum effects. Classically, one would expect all these correlation functions to be identical. 
However, the main finding of this study is summarized in Fig. \ref{f1}, displaying the computed pair correlation function $g(r)$ for pure metastable liquid \ph2, as well for a mixture with a \od2 concentration $X=0.49$. Also shown for comparison is the same quantity for solid (hcp) \ph2.  

The clear difference between the pair correlation function in the crystal and that of the other cases, shows that the simulation protocol adopted here is indeed effective in stabilizing a metastable fluid. But the most obvious observation is that within the statistical errors of the calculation there is no discernible difference between the pair correlation functions computed for \ph2 in the pristine fluid, or in the presence of \od2. In other words, the local environment experienced by \ph2 molecules is virtually unchanged by the presence of \od2. This is reflected in the kinetic energy per \ph2 molecule, which is identical in the pure fluid and in the mixture (close to 56 K), within statistical errors (of order 0.05 K).

Nothing structural, therefore, hints at any possible physical mechanism inhibiting crystallization in the mixture. Rather, this underscores once again the ``intermediate" character of the condensed phases of molecular hydrogen, on the one hand featuring significant quantum effects, on the other still largely behaving along classical lines. This is due to the strong suppression of quantum-mechanical exchanges, which underlie the most spectacular quantum manifestations, such as superfluidity \cite{feynman}, and also impart to the fluid phase of Bose systems its resilience against solidification at low temperature \cite{role}. The presence of \od2 (or any other foreign impurity) makes quantum exchanges, already virtually non-existent in pure fluid \ph2, even less likely; one would thus expect, if anything, the tendency of the system to crystallize to be strengthened in the mixture.

It should be mentioned that any cooling protocol, aimed at stabilizing on a computer a metastable phase, can only achieve  that goal for a finite time, ideally long enough to collect meaningful statistics. Eventually, the equilibrium (solid) phase inevitably must emerge, as long as the simulation methodology is ergodic, and reasonably efficient. Solidification has indeed been observed in this work, at the end of sufficiently long computer runs. It is noteworthy that no demixing of the two species is ever observed, i.e., the mixture crystallizes with the two species still mixed together, which is consistent both with the result illustrated in Fig. \ref{f1}, as well as with the general notion that quantum-mechanical exchanges also play a crucial role in the demixing of Bose mixtures \cite{PhysRevA.88.033628}. The observed resistance of the mixture to phase separation seems significant, as demixing is a possible mechanism that could inhibit crystallization.

In summary, we have carried out first principle QMC simulations of metastable liquid hydrogen mixtures, aimed at assessing whether crystallization can be inhibited in a mixture of parahydrogen and orthodeuterium. The results obtained indicate that the mixing of the two isotopic species does not result in a substantial change of the local environment experienced by \ph2 molecules that could in turn make the liquid phase more resilient than pure parahydrogen. No evidence of reduced molecular localization is seen, nor of isotopic demixing. Therefore, the results observed in this study suggest  that the actual reason for the experimental observation of the crystallization slowdown made in Ref. \cite{PhysRevB.89.180201} may warrant further investigation.

\section{Acknowledgments}
This work was supported by the Natural Sciences and Engineering Research Council of Canada.

 \bibliographystyle{elsarticle-num} 
 \bibliography{cas-refs}





\end{document}